%% file: main.tex
\documentclass{article}
\usepackage[english]{babel}
\usepackage{float}
\usepackage[letterpaper,top=2cm,bottom=2cm,left=3cm,right=3cm,marginparwidth=1.75cm]{geometry}
\usepackage{algorithmicx}
\usepackage{algpseudocode}
\usepackage{booktabs}
\usepackage{float}
\usepackage{algorithm}
\usepackage{amsmath}
\usepackage{graphicx}
\usepackage{longtable} 
\usepackage{array} 
\usepackage[colorlinks=true, allcolors=blue]{hyperref}

\usepackage{authblk}

\usepackage{titling}

\title{Best in Tau@LLMJudge: Criteria-Based Relevance Evaluation with Llama3}
\author{Naghmeh Farzi\\
  \texttt{Naghmeh.Farzi@unh.edu} \\
  University of New Hampshire \\
  Durham, New Hampshire, USA
  \and
  Laura Dietz\\
  \texttt{dietz@cs.unh.edu} \\
  University of New Hampshire \\
  Durham, New Hampshire, USA}

\date{}

\begin{document}

\maketitle









\begin{abstract}

Traditional evaluation of information retrieval (IR) systems relies on human-annotated relevance labels, which can be both biased and costly at scale. In this context, large language models (LLMs) offer an alternative by allowing us to directly prompt them to assign relevance labels for passages associated with each query. In this study, we explore alternative methods to directly prompt LLMs for assigned relevance labels, by exploring two hypotheses:

\textbf{Hypothesis 1} assumes that it is helpful to break down ``relevance'' into specific criteria---exactness, coverage, topicality, and contextual fit. We explore different approaches that prompt large language models (LLMs) to obtain criteria-level grades for all passages, and we consider various ways to aggregate criteria-level grades into a relevance label. \textbf{Hypothesis 2} assumes that differences in linguistic style between queries and passages may negatively impact the automatic relevance label prediction. We explore whether improvements can be achieved by first synthesizing a summary of the passage in the linguistic style of a query, and then using this summary in place of the passage to assess its relevance.

We include an empirical evaluation of our approaches based on data from the LLMJudge challenge run in Summer 2024, where our ``Four Prompts'' approach obtained the highest scores in Kendall's tau.
\end{abstract}

\section{Introduction}
The evaluation of information retrieval (IR) systems has historically depended on human-assigned relevance labels to assess the relevance of retrieved passages in response to different queries. However, acquiring large-scale relevance judgments from human annotators is costly and not free from biases. The emergence of large language models (LLMs) presents a scalable alternative, offering the potential for automated relevance assessments by directly prompting LLMs to evaluate and assign relevance labels \cite{Sun2023IsCG, MacAvaney2023OneShotLF, Thomas2023LargeLM, Faggioli2023PerspectivesOL}. While this approach facilitates labeling large datasets, it is met with skepticism regarding their reliability, primarily due to a lack of transparency in the underlying reasoning.


\bigskip
We study the task of automatically predicting passage-level relevance labels, to be used with a standard IR evaluation tool:

\textbf{Task Statement:} Given a query and a passage, predict the relevance on a graded scale. An example is the scale used in the LLMJudge \cite{rahmani_llmjudge_2024} challenge, which ranges from 0 to 3, where 3 indicates perfect relevance and 0 indicates no relevance.
\bigskip

In this study, we explore two key ideas for improving the relevance labeling process.

In \textbf{Hypothesis 1}, we assume that explicitly grading multiple criteria of relevance will help derive the final relevance label. 
We develop approaches by individually grading passages according to specific predefined relevance criteria that will be detailed in Section \ref{sec:hyp1}. To this end, we explore different prompt-based grading methods. Additionally, to aggregate the criteria-specific grades into the final relevance labels, we consider different approaches---ranging from prompt-based to using Naive Bayes aggregations.  In line with experience from Chain-of-Thought methods \cite{Wei2022ChainOT}, which gained from breaking down complex tasks, we believe this can enhance the evaluation quality---in addition to rendering the grading process more interpretable for a human that oversees the evaluation process.

With \textbf{Hypothesis 2} we assume that it is beneficial to substitute the passage with a more query-like representation. Hence we generate a synthesis of the passage text to be in line with the linguistic style of the query to create a more comparable basis in terms of presentation and content. The goal of this synthesis is to capture the overall topic and all relevant facets mentioned within the passage. We study whether bringing the passage into a question form will lead to a more accurate evaluation of relevance.

We empirically study the extent to which our approaches for assigning relevance labels enhance the evaluation of information retrieval systems, focusing on how well the resulting system rankings align with manual evaluations. We demonstrate a high inter-annotator agreement using Kendall's tau correlation coefficient \cite{Kendall1938ANM}. We also analyze the label-wise inter-annotator agreement between predicted relevance labels and manual judgments in terms of Cohen's Kappa and Krippendorff's Alpha \cite{457f3f87-2144-31f1-a31c-978d6b411185} measures. Additionally, the criteria-based grades provided by our Hypothesis 1 methods can offer valuable insights to human evaluators, helping them understand why a passage received a particular relevance judgment.

\paragraph{Outline.}
In this study, we develop five relevance labeling approaches to predict such labels for each query-passage pair, each approach is based on one of our two hypotheses. In Section \ref{sec:hyp1}, we will develop approaches based on Hypothesis 1, which involves grading passages based on predefined relevance criteria before deriving the relevance label. Section \ref{sec:hyp2} will cover an approach based on Hypothesis 2, where we propose representing passages in a way that aligns with the query's linguistic style before the comparison. Finally, in Section \ref{experimentaleval}, we present our experimental evaluation, comparing the effectiveness of these approaches by in terms of label-based inter-annotator agreement, and the correlation of the leaderboards, where different systems are ordered based on their evaluation score.

\section{Hypothesis 1 Approaches: Relevance Criteria} \label{sec:hyp1}

Hypothesis 1 posits that ``relevance'' can be broken down into four specific predefined criteria: exactness, coverage, topicality, and contextual fit, as elaborated below. We explore whether grading each criterion individually before deriving an overall relevance label can lead to improved evaluation. This methodology is based on two distinct phases: Criterion-Specific Grading and Relevance Label Prediction.

\begin{description}
    \item[Phase One: Criterion-Specific Grading:] Once the relevance criteria are defined in a structured set, each criterion is evaluated independently of the others for every passage-query pair, enabling a focused assessment of just one relevance criterion at a time. 

    \item[Phase Two: Relevance Label Prediction:] After grading the criteria, we aggregate the individual criterion grades from Phase One to produce a final relevance label that reflects the passage’s overall relevance in the context of the query. This aggregation allows for a more structured assessment of relevance.

\end{description}

A human overseeing this automatic process can verify each criterion grade as well as how it is aggregated to obtain a relevance label, providing an avenue for human interpretability in automatic grading.

\paragraph{Relevance Criteria.} In this study, we are using the following relevance criteria.  We will be using the criterion name and description as part of the prompt-based grading approach in Phase One as elaborated in Table \ref{tab:4prompts-phase1}. 

\begin{enumerate}
        \item 
        \begin{description}
            \item[Criterion Name:] Exactness 
            \item[Criterion Description:] How precisely does the passage answer the query.
            \item[Rationale:] Exactness requires that when a query presents a specific question or statement, the passage must provide a direct and accurate answer. This criterion emphasizes that the text should address the query with precise information, rather than offering a general explanation for the query. Ensuring exactness guarantees that the response effectively meets the user’s specific informational needs.
        \end{description}
        \item 
        \begin{description}
            \item[Criterion Name:] Coverage 
            \item[Criterion Description:] How much of the passage is dedicated to discussing the query and its related topics.
            \item[Rationale:]  A relevant passage should dedicate the majority of its content to the query and its related topics, ensuring a comprehensive and informative response.
        \end{description}        

        \item 
        \begin{description}
            \item[Criterion Name:] Topicality 
            \item[Criterion Description:] Is the passage about the same subject as the whole query (not only a single word of it).
            \item[Rationale:] The rationale is that the topic should closely align with the entire query. This criterion aims to prevent instances where a passage only addresses a single word or term from the query, as this may lead to an inadequate understanding of the query's specific subject matter and overall relevance.
        \end{description}        

        \item
        \begin{description}
            \item[Criterion Name:] Contextual Fit
            \item[Criterion Description:] Does the passage provide relevant background or context.
            \item[Rationale:]  This criterion evaluates whether the passage offers relevant background information that complements the overarching themes of the query. It emphasizes the importance of shared context in establishing coherence and relevance, distinguishing it from the narrower focus of topicality.
        \end{description}        
        
\end{enumerate}

To illustrate the application of these relevance criteria within the\textbf{ Four Prompts} method—the first approach based on Hypothesis 1, which will be explained next—Table \ref{tab:examples_4prompts} provides two examples demonstrating how the prompts are employed to assess relevance. The table shows the evaluation process for one query using two different passages, highlighting predicted grades from each phase of the method.

\input{tab-example-four-prompts}

\bigskip
In the following sections, we outline several variations of the approaches for implementing Phase One and Phase Two, detailing how each method structures the relevance assessment process. We implement four different approaches: the first three utilize the same Four Prompt methodology in Phase One for generating criterion-specific relevance grades but differ in the aggregation method used in Phase Two. The last approach incorporates a binary check before executing its two main phases.

\subsection{Four Prompts} \label{subsec:fourprompts}

The ``Four Prompts'' method involves evaluating the relevance of a passage to a query based on four specific predefined relevance criteria elaborated above: Exactness, Coverage, Topicality, and Contextual Fit. These are graded independently and then aggregated separately. Below, we explain each of the two phases involved in the grading process used in this approach:

\input{four-prompts-grading}

\textbf{Phase One: Criterion-Specific Grading:} For each of the predefined relevance criteria—Exactness, Coverage, Topicality, and Contextual Fit—we obtain grades by prompting the LLM to evaluate each criterion individually.  The ``Criterion-Specific Grading Prompt'' is detailed in Table \ref{tab:4prompts-phase1}, where the criterion name and description are inserted as previously defined, alongside the query and passage text.  These four criterion-specific prompts direct the LLM to evaluate and score the passage based on each criterion in accordance with the system message mentioned in Table \ref{tab:4prompts-phase1}.

\textbf{Phase Two: Prompt-Based Aggregation of Criterion Grades into The Relevance Label:} After obtaining the criterion-based grades in Phase One, this phase uses these grades to determine the overall relevance of the passage to the query and predict a single relevance label. An aggregator prompt, detailed in Table \ref{tab:4prompts-phase2}, instructs the LLM to assess the passage's overall relevance by integrating the criterion-specific grades along with the query and passage. This allows the LLM to generate a comprehensive relevance label based on all inputs, including the criterion grades from Phase One.





\eject{}

\begin{table}
\centering
\caption{Mapping sums of criterion grades to relevance labels.
\label{tab:score_to_grade}}
\begin{tabular}{cc}
\toprule
\textbf{Criterion Grade Sum} & \textbf{Relevance Label} \\
\midrule
10--12 & 3 \\
7--9   & 2 \\
5--6   & 1 \\
0--4   & 0 \\
\bottomrule
\end{tabular}

\end{table}
\subsection{Four Prompts + Summation Aggregation}
\label{subsec:agg-sum}
This method uses the same criterion-specific grades as the ``Four Prompts'' method but aggregates the grades differently to predict the relevance label.

\textbf{Phase One: Criterion-Specific Grading:} As in ``Four Prompts'' methods (Section \ref{subsec:fourprompts}).


\textbf{Phase Two: Summation-based Aggregation of Criterion Grades into The Relevance Label} This phase involves aggregating the criteria grades from Phase One, using a summation-based approach. The individual grades from each relevance criterion are summed to produce a total grade. To assign an overall relevance label to the passage, we establish specific thresholds for the total grade, determining the strictness or leniency of the relevance evaluation. These thresholds can be adjusted to align with the desired evaluation standards.

Table \ref{tab:score_to_grade}, presents a sample mapping between the total grades and relevance labels used in our experiments. This table illustrates how varying total grades are translated into specific relevance labels based on the defined thresholds.

\begin{table}
\centering
\caption{System messages and prompts used in Binary Check + Subset of Four Prompts approach. Top: prompt for the binary relevance check. Middle/Bottom: aggregation prompts for relevant/non-relevant decision branches in Algorithm \ref{algo:binary-check}.}
\begin{tabular}{p{14cm}}
\toprule
\textbf{Binary Relevance Checking Prompt} \\
\midrule
\begin{verbatim}Instruction: Given a passage and a query, predict whether the passage 
includes an answer to the query by producing either ``Yes'' or ``No''.
Question: {query} Passage: {passage} Answer: 
\end{verbatim} \\
\midrule
\textbf{Relevance Grading} \\
\midrule
System Message:
\begin{verbatim}You are a search quality rater evaluating the relevance of passages. Given a 
query and passage, you must provide a score on an integer scale of 2 or 3 
with the following meanings: 
2 = Highly relevant: The passage has some answer for the query, but the
answer may be a bit unclear, or hidden amongst extraneous information.
3 = Perfectly relevant: The passage is dedicated to the query and contains 
the exact answer.
\end{verbatim} \\
Prompt:
\begin{verbatim}The given passage is relevant to the query, please rate how relevant it is 
to the query. The output must be only a score (2 or 3) that indicates how 
relevant they are. 
Query: {Query} 
Passage: {Passage} 
Exactness: {Exactness grade} (From Phase One: Criterion-Specific Grading)
Coverage: {Coverage grade}(From Phase One: Criterion-Specific Grading)
Score:
\end{verbatim} \\
\midrule
\textbf{Non-Relevance Grading} \\
\midrule
System Message:
\begin{verbatim}You are a search quality rater evaluating the relevance of passages. Given a 
query and passage, you must provide a score on an integer scale of 0 or 1 
with the following meanings: 
0 = Irrelevant: The passage has nothing to do with the query. 
1 = Related: The passage seems related to the query but does not answer it. 
\end{verbatim} \\
Prompt:
\begin{verbatim}The given passage is irrelevant to the query, please rate how irrelevant it
is to the query. The output must be only a score (0 or 1) that indicates 
how irrelevant they are.
Query: {Query} 
Passage: {Passage} 
Topicality: {Topicality grade} (From Phase One: Criterion-Specific Grading)
Contextual Fit: {Contextual Fit grade}(From Phase One: Criterion-Specific Grading)
Score:
\end{verbatim} \\
\bottomrule
\end{tabular}
\label{tab:sun_decomposed}
\end{table}

\subsection{Four Prompts + Gaussian Naive Bayes Aggregation} 
\label{subsec:agg-naivebayes}


This method also uses the criterion-level grades from the ``Four Prompts'' method as inputs to a  Gaussian Naive Bayes model to predict the relevance label.


\textbf{Phase One: Criterion-Specific Grading:} As in ``Four Prompts'' methods (Section \ref{subsec:fourprompts}).

\textbf{Phase Two: Gaussian Naive Bayes Aggregation of Criterion Grades into The Relevance Label} In this phase, we take the criterion grades from Phase One to predict the passages' relevance labels. This is achieved by applying a Gaussian Naive Bayes model, implemented using the \texttt{sklearn.naive\_bayes} function from the Scikit-learn framework.\footnote{Available at \url{https://scikit-learn.org/stable/modules/naive_bayes.html}} This method involves training a Gaussian Naive Bayes classifier on the relevance criteria grades, allowing it to predict the final relevance label.


\begin{algorithm}
\caption{Relevance Labeling Process: Binary Check Followed by Criterion-Specific Evaluation. Prompt names are denoted in \textit{italics}.\label{algo:binary-check} }
\begin{algorithmic}[1]
\State \textbf{Input:} Query and Passage text
\State \textbf{Output:} Relevance Label

\If{\textit{Binary Relevance Check Prompt} (from Table \ref{tab:sun_decomposed}) result is ``Yes''}
    \State \textbf{Step 1:} Grade Exactness criterion
    \State Use \textit{Criterion-Specific Prompt} from Table \ref{tab:4prompts-phase1} to rate the Exactness criterion on a scale from 0 to 3.
    \State \textbf{Step 2:} Grade Coverage criterion
    \State Use \textit{Criterion-Specific Prompt} from Table \ref{tab:4prompts-phase1} to rate the Coverage criterion on a scale from 0 to 3.
    \State \textbf{Step 3:} Determine the Relevance Label
    \State Use the grades from Exactness and Coverage criteria and use \textit{Relevance Grading Prompt} in Table \ref{tab:sun_decomposed} to determine the final relevance label, which will be either 2 or 3.

\ElsIf{\textit{Binary Relevance Checking Prompt} result is ``No''}
    \State \textbf{Step 1:} Grade Contextual Fit criterion
    \State Use the \textit{Criterion-Specific Prompt} from Table \ref{tab:4prompts-phase1} to grade the Contextual Fit criterion on a scale from 0 to 3.
    \State \textbf{Step 2:} Grade Topicality criterion
    \State Use the \textit{Criterion-Specific Prompt} from Table \ref{tab:4prompts-phase1} to grade the Topicality criterion on a scale from 0 to 3.
    \State \textbf{Step 3:} Determine the Relevance Label
    \State Aggregate the grades from the Contextual Fit and Topicality criteria and use the \textit{Non-Relevance Grading Prompt} from Table \ref{tab:sun_decomposed} to determine the final relevance label, which will be either 0 or 1.

\EndIf
\\
\Return Relevance Label

\end{algorithmic}
\end{algorithm}

\subsection{Binary Check + Subset of Four Prompts}
\label{subsec:binary-followedby-criterion-specific}


In this approach, we hypothesize that different relevance criteria may carry varying levels of importance in different relevance regimes (lower relevance vs.\ high relevance). This method is based on the assumption that if we know whether a passage is more or less relevant, we can more effectively determine which criteria to prioritize in the relevance label. 

First, we use a direct relevance labeling prompt (here we use Sun's prompt \cite{Sun2023IsCG}) which acts as a binary check heuristic. This initial check categorizes the passage as either relevant or not relevant. Based on the results of this binary check, we proceed with a two-phased evaluation process. Phase One involves criterion-specific grading using a subset of relevance criteria informed by the binary check. Phase Two aggregates these criterion-specific grades to produce a final relevance label. Table \ref{tab:sun_decomposed} shows the prompts and system message in this process, and the detailed process provided in Algorithm \ref{algo:binary-check}.
\textbf{Phase Zero: Binary Relevance Check:} Obtain binary relevance by prompting the LLM to provide a simple ``Yes'' or ``No'' answer regarding the passage's relevance to the query. This relevance check allows us to later be able to differentiate passages of lesser from higher relevance.

\textbf{Phase One: Criterion-Specific Grading:}
In this phase, we evaluate the relevance of the passage based on a subset of relevance criteria, depending on the binary relevance response. We use the same relevance-criterion grading prompts as in the ``Four Prompts'' method (Section \ref{subsec:fourprompts}).


\begin{description}
\item [For a ``Yes'' response (indicating the passage is relevant):] The goal is to determine whether the relevance label should be 2 (Highly Relevant) or 3 (Perfectly Relevant). We prompt the LLM to grade the two criteria: Exactness and Coverage. We hypothesize these criteria effectively quantify the degree of relevance.

\item[For a ``No'' response (indicating the passage is non-relevant):] The goal is to determine whether the relevance label should be 0 (Irrelevant) or 1 (Related). In this case, the LLM grades the two other criteria: Contextual Fit and Topicality. We hypothesize that
these criteria help quantify the degree of non-relevance.
\end{description}

\textbf{Phase Two: Prompt-Based Aggregation of Criterion Grades into The Relevance Label:}  
In this phase, the final relevance label is determined based on the binary relevance output and grades from Phase One using a prompt-based aggregation.

For a ``Yes'' response (indicating relevance), we prompt the LLM to assign a label of either 2 or 3, guided by the ``Relevance Grading'' prompt and system message in Table \ref{tab:sun_decomposed}. This prompt uses the exactness and coverage grades obtained from Phase One, along with the query and passage. For a ``No'' response (indicating non-relevance), we prompt the LLM to assign a label of either 0 or 1 using the ``Non-Relevance Grading'' prompt and system message from Table \ref{tab:sun_decomposed}. This prompt utilizes the topicality and contextual fit grades obtained from Phase One, along with the query and passage.


\begin{table}
\centering
\caption{A One-shot prompt for evaluating relevance by synthesizing passage summaries into query style using Hypothesis 2.}
\label{tab:query_generation_prompts}
\begin{tabular}{p{14cm}}
\toprule
\textbf{Passage-to-Query Generation} \\ \midrule
System Message:
\begin{verbatim}
You are a query generator. For example, having this document:'Categories: Dogs. 
Article Summary X. If your puppy is starting to get teeth, it's probably between
3 and 4 weeks old. At 8 weeks of age, your puppy will have 28 baby teeth. 
For an adult dog, expect 1 or 2-year-olds to have white teeth, while 3-year-olds 
may have signs of tooth decay, such as yellow and brown tartar.' You should 
generate a query such as: 'dog age by teeth'.
\end{verbatim}\\
Prompt:
\begin{verbatim}
Please identify the search query that best corresponds to the following passage. Keep
your response concise. Passage: {passage}. 
\end{verbatim}
\\ \midrule
\textbf{Grading Text Similarity} \\ \midrule
System Message:
\begin{verbatim}
You are a similarity evaluator agent. Please rate the similarity between the two 
items on a scale from 0 to 3. 
\end{verbatim}
Prompt:
\begin{verbatim}
Please rate the similarity between the following queries: 
{Generated Query} 
and 
{Original Query} 
3: Highest similarity  
2: Fairly similar 
1: Minor similarity 
0: Not similar 
\end{verbatim}
\\ \bottomrule
\end{tabular}
\end{table}

\eject{}

\section{Hypothesis 2 Approaches: Passage-to-Query Generation} \label{sec:hyp2}

This approach explores the hypothesis that aligning the linguistic styles of passages and queries can improve relevance judgments. We posit that differences in linguistic styles between queries and passages can complicate comparisons, potentially impacting relevance labeling. To address this, we propose generating a query-like representation for each passage (which we call the generated query) that captures a potential query that this passage would answer.  We study whether the predicted relevance label can be improved by comparing the similarity between the original query and this generated query---instead of the original passage text.

\begin{table}
\caption{A worked example for Hypothesis 2. Generated queries are query-style representations of passages which are generated using the LLM. Similarity of generated queries and original query are evaluated using LLMs on a scale from 0 to 3, aligning with relevance labels. \label{tab:hypo2-example}}
\centering
\begin{tabular}{p{7cm} p{7cm}}
\toprule
\multicolumn{2}{c}{\textbf{Query ``q35''}: Do larger lobsters become tougher when cooked?} \\
\midrule
\textbf{Passage ``p8163''} & \textbf{Passage ``p4661''} \\
\midrule
I thought the whole bigger lobsters are tougher business was a myth. Larger lobsters are easier to overcook, making them tougher...but cooked properly they are no tougher. Also, meat from soft-shell lobsters is more tender than that from hard-shell lobsters. At least that's what I've read. & by the time a lobster gets to 3lbs, it is starting to get tough. long time cooking softens it up. a long time ago, we bought 6-8 lb lobsters in the waltham market. regular price \$0.79 a lb. special, \$0.69 a pound. those babies were tough. cook for about an hour, then chop them up for salad. Reply. slawecki.\\
\midrule
\textbf{Generated Query:} toughness of lobsters & \textbf{Generated Query:} cooking lobster \\
\textbf{Similarity Score Graded by Prompt:} 3 & \textbf{Similarity Score Graded by Prompt:} 2 \\
\midrule
\textbf{Ground Truth Relevance Label:} 3 & \textbf{Ground Truth Relevance Label:} 2 \\
\bottomrule
\end{tabular}
\end{table}



The first phase of this approach involves prompting the LLM to synthesize a query from a passage, using the prompt in Table \ref{tab:query_generation_prompts}, top. This generated query serves as a representative summary of the passage’s content, specifically tailored to align with the query's linguistic style and length.

In the second phase, we prompt the LLM to evaluate the similarity between the generated query and the original query using a scale from 0 to 3, which directly corresponds to our relevance labeling system. Higher similarity scores indicate a closer alignment between the passage’s content and the query’s intent. For this assessment, we use the prompt in Table \ref{tab:query_generation_prompts}, bottom.

Table \ref{tab:hypo2-example} provides a worked example for this approach, illustrating how query generation and similarity scoring were applied to query ``q35''. The example shows how the queries that are generated from the passage represent its core theme while following the linguistic style of keyword queries. 

Comparing the generated queries ``toughness of lobsters'' and ``cooking lobsters'' to the given query ``Do large lobsters become tougher when cooked?'' with the similarity prompt, allows the LLM to place a preference on the toughness aspects, as it is more specific to the query.



\eject{}
\section{Experimental Evaluation} \label{experimentaleval}

In this section, we present the experimental evaluation of our approaches based on \textbf{Hypothesis 1}, which involves grading predefined criteria prior to assigning relevance labels, and \textbf{Hypothesis 2}, which focuses on representing the passage in a query-like form to achieve closer linguistic alignment with the original query prior to comparison. 

\subsection{Experimental Setup}

\paragraph{Dataset.}
To assess the effectiveness of our methods, we participated in the LLMJudge challenge as part of the LLM4Eval workshop at SIGIR 2024. The challenge is based on queries and run files of the TREC Deep Learning 2023 dataset. The organizers collected a set of systems to be evaluated\footnote{At the time of writing, the systems (or run files) were not shared with participants.  Due to the absence of the system run files, we are unable to reproduce the results directly; however, the insights gained from the reported metrics allow us to draw meaningful comparisons.  }, obtained judgment pools, and asked challenge participants to predict relevance labels for each passage/query in the pool.

The dataset is built upon the passage retrieval task dataset of the TREC 2023 Deep Learning track\footnote{Data set available at \url{https://microsoft.github.io/msmarco/TREC-Deep-Learning.html}} (TREC-DL 2023) \cite{craswell2023} and includes 25 test queries and 25 development queries, with a total of 7,224 passages for the test queries and 4,414 passages for the development queries. For both sets, there is an average of approximately 81 relevant passages (including both relevant and highly relevant) per query. The relevance grading scale follows that of the TREC Deep Learning track, where grades 0, 1 indicate non-relevant passages, and 2, 3 relevant passages (higher = more relevant).

\paragraph{Model setup.}

All of our methods use an out-of-the-box \texttt{Meta-Llama-3-8B-Instruct} language model, which comprises 8 billion parameters, deployed on NVIDIA A40 hardware. While any LLM could be used here, we find that this model is particularly well-suited for grading according to different criteria.

\paragraph{Handling of invalid prompt responses.}
The current implementation employs error-handling strategies to manage various unwanted situations during the evaluation process.

When a CUDA out-of-memory error is detected—often due to a passage exceeding the length limit or not being properly cleaned—the script logs the error, skips the problematic passage, and assigns a label of 0. This prevents the evaluation process from halting due to memory constraints and ensures that the process continues smoothly.

Similarly, if the predicted label is not a valid integer, the script logs the error and assigns a label of 0 to maintain consistency and avoid invalid entries in the results.

The script also manages responses from the language model to extract a valid relevance label. Occasionally, the model may return a full sentence explaining the answer alongside the numeric relevance label. To handle this, the script parses the response and extracts the first standalone number within the range of 0 to 3. This ensures that the final label conforms to the expected grading scale, maintaining accuracy and consistency in the evaluation results. If the label is not acceptable or falls outside the expected range, a label of 0 is logged for that query-passage pair.

\paragraph{Evaluation measures.}

We focus on three evaluation measures provided by challenge organizers: leaderboard correlation measured by Kendall's Tau, and inter-annotator agreement measures by  Krippendorff's Alpha and Cohen's Kappa.

Focusing on the evaluation of submitted systems, our main metric is Kendall's tau, which measures the rank correlation of systems on the leaderboard based on manual assessments (ground truth) and the predicted system evaluation scores using the relevance label of each method. Kendall’s Tau assesses the ordinal relationship between leaderboard rankings, ranging from -1 to +1, where +1 indicates perfect agreement and 0 indicates no agreement. This measure is crucial for evaluating how well different evaluation methods rank the quality of retrieval systems. This aligns without the goal of differentiating retrieval systems by their quality, placing the best systems on the top of the leaderboard, and better-performing systems above inferior systems.

Among all submissions to the LLMJudge challenge, our ``Four Prompts'' method offered the best rank correlation with the manual leaderboard.

We additionally report inter-annotator metrics, such as Cohen's Kappa and Krippendorff’s Alpha, that measure how well the predicted relevance labels align with manually assessed labels. 
While high inter-annotator agreement implies consistency among evaluations, discussions during the LLM4Eval workshop underscored the priority of Kendall’s Tau for our objectives. This measure specifically focuses on the correlation of leaderboards, allowing us to identify the best-performing systems more effectively.

\paragraph{Compared evaluation methods and baselines.}

In this study, we focus on the comparison of the approaches described in Table \ref{tab:approaches_summary}. For reference, we also include the methods that obtained the best tau (\textcolor{blue}{*}), kappa ({\textcolor{red}{*}}/ {\textcolor{yellow}{*}}), and alpha ({\textcolor{green}{*}}) among all submissions in the LLMJudge challenge in Table \ref{tab:testresult}.

\input{table-approach-identifiers}

\begin{table}
\caption{Comparison on the test set of the LLMJudge/TREC DL2023 dataset, as reported at the LLM4Eval workshop. We denote the most successful approaches in the LLMJudge challenge as follows: best in rank correlation, as measured in Kendall's Tau (\raisebox{-0.5ex}{\textcolor{blue}{*}}).  Best in inter-annotator agreement, measured in 4-point Cohen's Kappa (\raisebox{-0.5ex}{\textcolor{red}{*}}),   01 vs 23 Cohen's Kappa (\raisebox{-0.5ex}{\textcolor{yellow}{*}}), or Krippendorff's Alpha (\raisebox{-0.5ex} {\textcolor{green}{*}}).  
\\
Among approaches described in this paper, the best method is denoted in bold; the best among all submissions to the challenge is denoted in bold italics.}
\centering
\resizebox{\textwidth}{!}{%
\begin{tabular}{p{5cm}cccccc} 
\toprule
 \textbf{Identifier} & \textbf{Tau} & \textbf{Alpha} & \multicolumn{4}{c}{\textbf{Kappa}} \\
\cmidrule{4-7}
   & &  &  \textbf{4-point} & \textbf{0 vs 123} & \textbf{01 vs 23} & \textbf{ 012 vs 3}  \\
\midrule
 \textbf{TREMA-4prompts}\raisebox{-0.5ex} {\textcolor{blue}{*}} &\textbf{\textit{0.9483}} & 0.2888 & 0.1829 & 0.3022 & 0.2697 & 0.1664 \\
 TREMA-sumdecompose & 0.9300 & \textbf{0.3926} & \textbf{0.2088} & \textbf{0.3228} & \textbf{0.3512} & \textbf{0.2047}  \\
 TREMA-naiveBdecompose & 0.9128 & 0.3579 & 0.1741 & 0.3085 & 0.2916 & 0.0153 \\
TREMA-CoT & 0.8956 & 0.3852 & 0.1961 & 0.3181 & 0.3208 & 0.1836 \\
TREMA-other & 0.8200 & 0.2712 & 0.1408 & 0.2740 & 0.2015 & 0.1411 \\
\hline
\textbf{willia-umbrela1} \raisebox{-0.5ex}{\textcolor{red}{*}} & 0.9009 & 0.4918 & \textbf{\textit{0.2863}} & 0.4161 & 0.3985 & 0.3145 \\
\textbf{h2oloo-fewself} \raisebox{-0.5ex}{\textcolor{yellow}{*}} & 0.9085 & 0.4958 & 0.2774 & 0.4172 & \textbf{\textit{0.4280}} &0.3048\\
\textbf{Olz-gpt4o} \raisebox{-0.5ex} {\textcolor{green}{*}} & 0.8793 & \textbf{\textit{0.5020}} & 0.2625 & 0.4228 & 0.3657 & 0.3066\\
\bottomrule
\end{tabular}%
}

\label{tab:testresult}
\end{table}

\subsection{Results}

\paragraph{Leaderboard Correlation (Kendall's Tau).}
Our findings reveal that the ``Four Prompts'' method 
 yields the best Kendall's Tau among all approaches in the LLMJudge challenge. This highlights the performance of using our method for the purpose of evaluating different information retrieval systems to study their respective quality in comparison.
 
 The superior performance of the ``Four Prompts'' method underscores that breaking down relevance into predefined criteria and using specific criterion grades to assign relevance labels indeed leads to improved evaluation performance.

\paragraph{Inter-annotator Agreement.}
We focus on leaderboard correlation measures like Kendall's tau, as these are the most important for the purpose of evaluating different systems. However, we acknowledge that LLM-based relevance labeling approaches may also be used to generate synthetic training data. For the purpose of generating training data, it is more important to obtain a high inter-annotator agreement (as measured with Cohen's Kappa and Krippendorff's Alpha). 


The best method in terms of Krippendorff's Alpha is ``Olz-gpt4o'' the best method in terms of 4-point Cohen's Kappa is ``willia-umbrela1'' and the best method in terms of 01 vs.\ 23 Cohen's Kappa is ``h2oloo-fewself''. These methods obtain a noticeably lower leaderboard correlation. 

\paragraph{Jack-of-all-Trades.}
Among our submitted approaches, the ``TREMA-sumdecompose'' method obtains a higher inter-annotator score than ``TREMA-4prompts'', both in terms of Krippendorff’s alpha and different Cohen's Kappa, while still outperforming ``Olz-gpt4o'', ``willia-umbrela1'', and ``h2oloo-fewself'' in terms of Tau. This demonstrates that the criterion-based approach can yield strong agreement with manual relevance judgments when the aggregation method is chosen differently---providing a good balance between leaderboard correlation and inter-annotator agreement.



\section*{Acknowledgment}
We would like to express our gratitude to the organizers of the LLM4Eval workshop for providing a platform to present our work. We appreciate all the feedback and discussion received from other participants.

\bibliographystyle{plain}
\bibliography{sample}

\end{document}

%% file: tab-example-four-prompts.tex
\begin{table}
\centering
\caption{Examples of prompts and corresponding scores for query ``q18'' with two passages. This table illustrates the evaluation process in Phase One, showcasing the criterion scores for Exactness, Coverage, Topicality, and Contextual Fit for each passage. The Phase Two relevance labels are derived from these scores, allowing for a comparison with the ground truth relevance labels.
\label{tab:examples_4prompts}}
\begin{tabular}{p{7cm} p{7cm}}
\toprule
\multicolumn{2}{c}{\textbf{Query ``q18''}: dog age by teeth} \\
\midrule
\textbf{Passage ``p4068''} & \textbf{Passage ``p75''} \\
\midrule
Puppies start to get their puppy teeth at the age of 3 to 4 weeks. They will start with 28 puppy teeth. These teeth will be replaced with their 42 permanent adult teeth at about the age of four months. Dogs have four different types of teeth & Humans and most other mammals have a temporary set of teeth, the deciduous, or milk, teeth; in humans, they usually erupt between the 6th and 24th months. These number 20 in all: 2 central incisors, 2 lateral incisors, 2 canines, and 4 premolars in each jaw. At about six years of age, the preliminary teeth begin to be shed as the permanent set replaces them.\\
\midrule
\textbf{Phase 1 - Criterion Scores} & \textbf{Phase 1 - Criterion Scores} \\
\midrule
\textbf{Exactness:} 2 & \textbf{Exactness:} 0 \\
\textbf{Coverage:} 2 & \textbf{Coverage:} 0 \\
\textbf{Topicality:} 3 & \textbf{Topicality:} 0 \\
\textbf{Contextual Fit:} 3 & \textbf{Contextual Fit:} 0 \\
\midrule
\textbf{Phase 2 - Relevance Label:} 2 & \textbf{Phase 2 - Relevance Label:} 0 \\
\midrule
\textbf{Ground Truth Relevance Label:} 2 & \textbf{Ground Truth Relevance Label:} 0 \\
\bottomrule
\end{tabular}
\end{table}

%% file: four-prompts-grading.tex
\begin{table}

\centering
\caption{System messages and prompts for criterion grading for the ``Four Prompts'' method (and other approaches based on Hypothesis 1).\label{tab:4prompts-phase1}}
\begin{tabular}{p{14cm}}
\toprule
\textbf{Criterion-Specific Grading} \\ 
\midrule
System Message:
\begin{verbatim}
Please assess how well the provided passage meets specific criteria in
relation to the query. Use the following scoring scale (0-3) for evaluation:

0: Not relevant at all / No information provided.
1: Marginally relevant / Partially addresses the criterion.
2: Fairly relevant / Adequately addresses the criterion.
3: Highly relevant / Fully satisfies the criterion.
\end{verbatim} \\
Prompt:
\begin{verbatim}
Please rate how well the given passage meets the {Criterion Name} criterion in
relation to the query. The output should be a single score (0-3) indicating
{Criterion Description}.
Query: {Query}
Passage: {Passage}
Score:
\end{verbatim} \\
\bottomrule
\end{tabular}
\end{table}

\begin{table}
\centering
\caption{System messages and prompts for aggregating criterion-level grades (Phase Two) of the Four Prompts method: Aggregating scores to generate comprehensive relevance labels for each passage and query.
\label{tab:4prompts-phase2}}
\begin{tabular}{p{14cm}}
\toprule
\textbf{Aggregating Criterion Grades into a Relevance Label} \\
\midrule
System Message:
\begin{verbatim}
You are a search quality rater evaluating the relevance of passages. Given a 
query and passage, you must provide a score on an integer scale of 0 to 3
with the following meanings: 
3 = Perfectly relevant: The passage is dedicated to the query and contains
the exact answer. 
2 = Highly relevant: The passage has some answer for the query, but the
answer may be a bit unclear, or hidden amongst extraneous information. 
1 = Related: The passage seems related to the query but does not answer it. 
0 = Irrelevant: The passage has nothing to do with the query. 
Assume that you are writing an answer to the query. If the passage seems to
be related to the query but does not include any answer to the query, mark
it 1. If you would use any of the information contained in the passage in 
such an answer, mark it 2. If the passage is primarily about the query, or 
contains vital information about the topic, mark it 3. Otherwise, mark it 0.  
\end{verbatim} \\
Prompt:
\begin{verbatim}
Please rate how the given passage is relevant to the query based on the
given scores. The output must be only a score that indicates how relevant
they are. 
Query: {Query} 
Passage: {Passage} 
Exactness: {Exactness Score} (From Phase One)
Topicality: {Topicality Score} (From Phase One)
Coverage: {Coverage Score} (From Phase One)
Contextual Fit: {Contextual Fit Score} (From Phase One)
Score:
\end{verbatim} \\
\bottomrule
\end{tabular}

\end{table}

%% file: table-approach-identifiers.tex
\begin{table}
\centering
\caption{Overview of compared approaches. We use identifiers submitted to the LLMJudge challenge, detailing which methods they represent, and how they differ from one another.}
\begin{tabular}{lcccc}
\toprule
\textbf{Identifier }& \textbf{Phase Zero}   & \textbf{Phase One }     & \textbf{Phase Two} &
\textbf{Sect.}\\
\midrule
\textbf{Hypothesis 1} & \textbf{Pre-check} & \textbf{Grading} & \textbf{Aggregation}\\
TREMA-4prompts   & -   & Four Prompts    & Prompt-based  & \ref{subsec:fourprompts}     \\
TREMA-sumdecompose & -   & Four Prompts   & Summation & \ref{subsec:agg-sum}\\

TREMA-naiveBdecompose & -   & Four Prompts   & Naive Bayes & \ref{subsec:agg-naivebayes}\\
TREMA-CoT & Binary Check & Subset of Four Prompts & Prompt-based   & \ref{subsec:binary-followedby-criterion-specific}    \\
              \midrule
\textbf{Hypothesis 2}  & - & \textbf{Representation} & \textbf{Matching} \\
TREMA-other & -            & Query Generation      & Similarity Check  &\ref{sec:hyp2}    \\
\bottomrule   
            
\end{tabular}
\label{tab:approaches_summary}
\end{table}